Detergents and chaotropes for protein solubilization before two-dimensional electrophoresis

Thierry RABILLOUD*,
* : Laboratoire, d'immunologie, DRDC/ICH, INSERM U 548, CEA Grenoble, 17, rue des martyrs, F38054 GRENOBLE CEDEX 9 FRANCE

Abbreviations
ASB14: 3-(tetradecanoylamidopropyl dimethylammonio) propane 1-sulfonate
C7BzO: 3-(4-Heptyl)phenyl-3-hydroxypropyl)dimethylammoniopropanesulfonate
C13E10 deca(ethylene oxide) tridecyl ether
DTT: dithiothreitol
CA-IEF: carrier ampholyte isoelectric focusing
IEF: isoelectrofocalisation
IPG : immobilized pH gradient
pI: isoelectric point
TBP : tri-butyl phosphine
TCEP: tris-carboxyethyl phosphine
SDS-PAGE: sodium dodecyl sulfate-polyacrylamide gel electrophoresis
2D: two-dimensional

Abstract
Because of the outstanding separating capabilities of two-dimensional electrophoresis for complete proteins, it would be advantageous to be able to apply it to all types of proteins, including membrane proteins. Unfortunately, severe solubility problems hamper the analysis of this class of proteins. As these problems arise mainly in the extraction and isoelectric focusing steps, solutions are sought to improve protein solubility under the conditions prevailing during isoelectric focusing. These solutions deal mainly with chaotropes and new detergents, which are both able to enhance protein solubility. The input of these compounds in proteomics analysis of membrane proteins is discussed, and future directions are also discussed

# 1. Introduction

The solubilization of proteins for 2D electrophoresis-based proteomics is a difficult task. 2D electrophoresis with IEF in the first dimentsion is almost one of the worst possible setups for protein solubilization. Proteins must reach their pI, which is also their minimum of solubility, and still stay soluble at the pI. Moreover, as the protein mobility decreases when proteins get close to their pI, IEF is performed under strong electric fields (200V/cm compared to the 10-20V/cm used for SDS-PAGE). This means in turn that salts and ionic compounds in general are almost banned in IEF. Moreover, any solubilizing agent used prior to IEF must not change the original pI of the proteins. Consequently, this precludes the use of strong ionic detergents (e.g.SDS). However, low amounts (up to 0.03% w/v) of ionic detergents can be used, provided that conditions favoring the exchange of SDS for other, non-ionic detergents are used in IEF [1-3]. This ensures removal of bound SDS from the proteins, but this also means that the benefits of SDS are lost for the IEF dimension. However, the use of SDS has been often recommended as a way to ensure a complete initial solubilization before IEF. Apart from these problems arising from the protein world, other problems are frequently encountered in many biological samples. These problems arise form the non-proteinaceous compounds that can be present in the sample. A canonical example is the one of nucleic acids, which completely blur the 2D electrophoresis pattern when present at too high a concentration [4]. Nucleic acids act as mobile ion exchangers at the low ionic strength required by IEF, thereby creating severe artefacts. Other classes of compounds (lipids, salts etc.) can be encountered in many samples and create their artefacts. .

There are thus different problems depending on the starting material. This chapter will however focus on one side of the problem; i.e. the intrinsic solubilization of proteins, and will mainly deal with the chaotropes and detergents used for initial solubilization and for IEF. There is an intricate play between chaotropes and detergents for protein solubilization before IEF and 2D electrophoresis. As a matter of facts, both are necessary under these conditions, as shown by the fact that systems running with detergents alone or chaotropes alone are very poorly efficient, opposite to the efficiency of SDS in SDS-PAGE. Starting from this example, it is quite clear that non-ionic detergents are not able to denature and solubilize proteins per se as SDS is able to do. This is precisely due to the fact that detergents compatible with IEF do not bear any electrical charge, while this feature is essential to the efficiency of ionic detergents such as SDS, bur also cationic detergents such as CTAB or BAC16 that are use in other setups of 2D electrophoresis (see other chapters of this book). Because of this poor efficiency of electrically neutral detergents, it is necessary to add chaotropes for a better denaturation and solubilization of proteins.

The role of chaotropes inthe solubilization process is to break the inter-and intra-molecular non-covalent interactions in the sample (e.g. hydrogen bonds, dipole-dipole interactions, and hydrophobic interactions), and thereby facilitate protein unfolding. Although ionic bonds are not directly affected by nonionic chaotropes such as urea and thiourea, the influence of these chaotropes on the dielectric constant of water also alters the strength of the ionic bonds. However, chaotropes alone are not efficient for complex membrane samples. For example, they are unable to solubilize the lipid bilayer of membranes, which is required for membrane protein solubilization. Futhermore, by unfolding proteins, chaotropes increase the number of hydrophobic residues exposed to the solvent, and thus the possibilities of protein aggregation. The addition of detergents, which are specialized molecules for handling hydrophobic effects in a water-based solvent, limits considerably this aggregation problem.

As mentioned, the constraints imposed by IEF preclude the use of substances which are ionized within the pH range used for IEF, that is only nonionic or zwitterionic compounds can be used. This narrows the choice for chaotropes to the amide and urea families, as guanidines and amidines are charged below pH 12. Among the possible chaotropes, urea has been used for quite a long time. More recently, the addition of thiourea to urea as an additional chaotrope has shown interesting features for protein solubilization [5] but also to limit protease action [6].

Among the wide choice of commercially available uncharged detergents, two subfamilies can be distinguished. Nonionic detergents do not have any charges on the molecule, while zwitterionic detergents have an equal number of negative and positive charges on their molecules. Depending on the pK of the ionizable groups on the molecules, some detergents can be ionic in a certain pH range (where at least one group is titrated) and zwitterionic in another pH range, while other detergents can be zwitterionic on the complete pH range. As an example of the two cases, classical betaines (bearing a quaternary ammonium and a carboxylic group) are positively charged at low pH, when the carboxylic group is not fully deprotonated. When the carboxylic group is fully deprotonated, i.e. more than 2 pH units above the pK, they behave as a zwitterionic detergent. Conversely, sulfobetaines, bearing a quaternary ammonium and a sulfonic group, are zwitterionic over the 0-14 pH range, as both groups are ionized in this range. Of course, only detergents completely zwitterionic over the pH range of interest can be used for IEF.

The detergents historically used for 2D electrophoresis are Triton X100 (or NP-40) (nonionic) and CHAPS (zwitterionic). These have been used extensively in combination with urea, and have not proved very efficient for the solubilization of sparingly soluble proteins, e.g. membrane proteins [7]. However, recent work has shown that either specially-designed zwitterionic detergents [8-10], or carefully-selected non ionic detergents [11] can solubilize membrane proteins. It is interesting to note that Triton X100 is poorly efficient when used with urea alone and much more efficient in urea-thiourea [11]. This is also true of linear alkyl oligo ethylene glycol compounds [11], such as the Brij ® detergents . However, the most efficient nonionic detergents belong to the glycoside family (e.g. octyl glucoside, dodecylmaltoside), and the latter seems to be efficient in urea alone [12] as well as in urea-thiourea [11, 13]. An example of the variations in protein solubilization induced by the choice of detergent can be seen in figure 1. The multiple variables playing a role in the solubilization process have also been investigated in [14]. However, it should not be concluded from the preceding that dodecyl maltoside is the absolute best choice for protein solubilization for 2D electrophoresis. The optimal detergent will depend on the sample. However, there is some kind of a shortlist, based on previous work. So the best candidates for protein solubilization, at least as a first screen, are to be chosen among dodecyl maltoside, ASB14, C7BzO, Brij56 and C13E10.

## 2. Materials

### 2.1. Biological material

Human red blood cell membranes, which are a very good control for membrane protein solubilization tests, are prepared as follows (all steps are carried out between 0 and 4°C):
10 ml of freshly drawn human blood, collected on EDTA or citrate, are diluted with 10ml of PBS supplemented with 1mM EDTA. The diluted blood is loaded on a 10ml cushion of Ficoll/sodium diatrizoate solution (d:1.077, trade names Ficoll Hypaque or equivalent) and this is centrifuged at 2000g for 20 minutes. the supernatant (plasma and white blood cells) is eliminated, and the red blood cells is resuspended in 10 ml of PBS-EDTA and recentrifuged at 2000g for 20 minutes. The pellet is saved, and suspended in 20 times its volume of 2mM EDTA pH 8, resulting in hemolysis. The hemolysate is centrigued at 50000g for 20 minutes. The dark red supernatant is eliminated, leaving a double-layered pellet, with a tigh, light red pellet at the bottom and a bluish-white, more fluffy pellet on top. This white layer is the membrane pellet, the red layer being composed of protease-rich granules. The white layer is resuspended in 30 ml PBS-EDTA, recentrifuged at 50000g for 20 minutes, and the supernatant is removed. One more washing is made with 2mM EDTA. The final pellet is resuspended in a minimal volume of Tris-HCl 10mM pH 7.5, sucrose 0.2M, EDTA 1mM, aliquoted and stored at -80°C (preferably) or -20°C (more limited shelflife). The protein concentration is estimated by a standard protein assay.

2.2. Equipment

1. A tabletop ultracentrifuge, used for membrane preparation and to remove unsolubilized proteins
2. Immobilised pH gradient (IPG, linear and non-linear pH gradient from 3 to 10, 18 cm length, Amersham Pharmacia Biotech).
3. IPGphor apparatus: for isoelectrofocalisation of proteins (Amersham Pharmacia Biotech).
4. Tube gels electrophoresis Setup (BioRad), for first dimension gel electrophoresis
5.Protean II: for SDS-PAGE electrophoresis (Biorad).

2.3. Reagents

2.3.1. Products and stock solutions

1. Dodecyl maltoside, Triton X100, and CHAPS are best used from 20% (w/v) stock solutions in water. These solutions should be stored at 4°C and show limited conservation (a few weeks)
2. C13E10, Brij 56, ASB14 are best used from 20% (w/v) stock solutions in ethanol/water (50/50 v/v). These solutions are stable for months at room temperature
3. Urea stock solution for IEF. It is difficult to go beyond 9M urea at room temperature, which is the concentration used when urea is the sole chaotrope. This means that urea is added as a solid (Note 1)
4. Urea-thiourea stock solution. The final chaotrope concentrations are 7M urea and 2M thiourea. This means that a 1,25x concentrated solution can be prepared, which is then simpler to use that reweighing small amounts of solid urea and thiourea for each sample. For 10 ml of this concentrated solution, weigh 5.25g of urea and 1.9g of thiourea. Some detergents (e.g. CHAPS or Triton X-100) which are fully compatible with urea, can be added at this stage. Other detergents, which show a more limited urea compatibility (e.g. ASB14), must be added only when the solution is diluted to the final strength. A total volume of 4.2 ml of liquid must be added to the urea and thiourea to make up for 10 ml (see also Note 2 and Note 3). This solution is stable for months if stored frozen at -20°C.
5. Tributylphosphine is a liquid (4M when pure). A 40 fold dilution in dimethyl formamide is made just prior to use. This solution is further diluted 50 fold in the sample solution.
6. Tris carboxyethyl phosphine is a solid. A 1M stock solution in water is made, and stable for months at -20°C

3. Methods (see Note 4)

3.1 Solubilization in urea for IEF

3.1.1. Solubilization from a solid sample, e.g. tissue or cell pellet
In this case, the sample volume can often be neglected in the final solubilization volume. A sample solution containing urea (9 to 9.5 M final concentration, see Note 1), the selected detergent (taken in the list above in 2.3.1) at 2 to 4% w/v concentration, carrier ampholytes (0.4% w/v for IPG, 2% w/v for CA-IEF) and a reducing agent (50mM DTT or 5mM TBP or 5mM TCEP). This solution is added to the solid sample, resulting in a liquid extract. Protein extraction is helped by sonication in a water bath sonicator for ca. 30 minutes. Unsolubilized material is best removed by ultracentrifugation for 30 minutes at 200,000g at room temperature.

3.1.2. Solubilization from a suspension or solution
In this case, the volume of the sample must generally be taken into account. It is thus necessary to calculate the final solution. As a rule of thumb, the sample volume can represent up to 35% of the final extraction volume. Solid urea, water and stock solutions of the detergent, ampholytes and reducer are used in addition to the liquid sample to build the extraction solution.

3.2. Solubilization in urea-thiourea for IEF
With the spreading of IPG with sample application by in-gel rehydration [15], rather large sample volumes can be used. This is especially true when home made strips are used, as these can be made wider than commercial IPG strips and thus accommodate a larger volume (up to 1 ml). It is thus often possible to use a dilution approach with the concentrated chaotrope solution and the solid sample resuspended in a minimal volume of water or the liquid sample. If the detergent can be predissolved in the concentrated chaotrope solution, which can also contain the reducer, then the sample volume can represent up to 20% of the total extraction volume. If the detergent must be added only at last, with a urea concentration not exceeding 8M, then it is more convenient to use 1 volume of sample, 1 volume of detergent stock solution, and to add 8 volumes of concentrated chaotrope solution. If this approach leads to too high a volume, two alternate approaches can be considered:
i) introduce in a sample tube a volume of a stock detergent solution equal to the sample volume. Evaporate the solvent in a SpeedVac. Add the sample and 4 volumes of concentrated chaotrope solution.
ii) Consider the sample volume will make 40% of the final sample volume. Weigh the corresponding amounts of urea, thiourea and solid detergent. Dissolve with the sample in a bath sonicator.
In all cases, an extraction time of 30-60 minutes at room temperature is optimal before centrifugation (200,000g, 30 minutes, room temperature) to remove unsolubilized material.

3.3. Running IEF gels with membrane proteins
The detergents used historically for IEF and 2D electophoresis of soluble proteins (e.g. CHAPS or Triton X100) are fully compatible with urea (see Note 5). This means in turn that the temperature control is needed just to keep urea and thiourea in solution, i.e. a temperature greater than 15°C.
However, some of the detergents used for better solubilization of membrane proteins, and especially those with a linear alkyl tail (e.g. ASB 14, ASB 16, Brij56) show a urea tolerance that is strongly temperature dependent (see Note 5). Furthermore, detergent precipitation just ruins the separation process. Consequently, the detergent concentration and running temperature must be carefully controlled. For all detergents mentioned here, using 2% detergent in a 7M urea/2M thiourea solution and running the gels at 22°C (thermostated) ensures complete solubility of the chaotrope-detergent mixture and an optimal solubilization power, in the sense that further increase in detergent concentration does not make new proteins to appear on the 2D gels.

Downstream steps of 2D electrophoresis (IEF gel equilibration, SDS-PAGE) do not need any adjustment compared to standard protocols for soluble proteins

## 4. Notes

1. The partial specific volume of urea is a useful number to know to make concentrated urea solution. 1 g of urea occupies 0.75 ml in solution. In the same order for most detergents, 1g occupies 1 ml. This is also true for thiourea. Urea must also never be warmed above 37°C to limit protein carbamylation. As an example, 1 ml of aqueous extract is added to 900 mg urea. this results in a 1.675 ml of a 9M urea solution. Otherwise, 1g urea is added to 1 ml aqueous extract, resulting in 1.75 ml of a 9.5M urea solution.

2. As a matter of facts, most of these extraction solution contain less than 50% water. This means that dissolution of the solids is rather difficult to perform, especially because high temperatures cannot be used (see previous note). The use of a water bath sonicator (marketed for cleaning objects and glassware) is of great help for this difficult solubilizations.

3. many detergents are not fully compatible with urea. depending on the detergent structure, on the urea concentration and on the temperature, insoluble detergent-urea complexes can form. Detergents with linear alkyl chain are especially prone to this problem (e.g. ASB 14, Brij 56) which completely prevents the use of commercial linear sulfobetaines, which do not stand more than 4M urea.

4. Many detergents strongly interfere with some popular protein assay methods, while others are plagued by reducing compounds . It is therefore recommended to take this dimension into account when performing protein solubilization. In some cases, one can end with a solubilization cocktail which is incompatible with any protein assay method. In this case, it is often advisable to determine the protein concentration in the initial sample, prior to extraction, especially if the sample is a suspension. This means in turn that it will not be possible to assess the efficiency of the solubilization process.

5. When used at high concentrations, urea forms channel-like structures in water. These channels are able to host organic compounds, especially those with a linear, saturated alkyl chain, to form so-called inclusion compounds that are much less soluble than either component taken alone. This is the molecular basis for the urea-driven precipitation of many detergents, which is also highly temperature-dependent, as is the solubility of the detergents-urea inclusion compounds.

This problem can be alleviated by subtle changes in the molecular structure of the detergents, e.g. inclusion of an amide bridge or using mixed alkylaryl hydrophobic parts, as aryl structures are too big to insert into the urea channels.

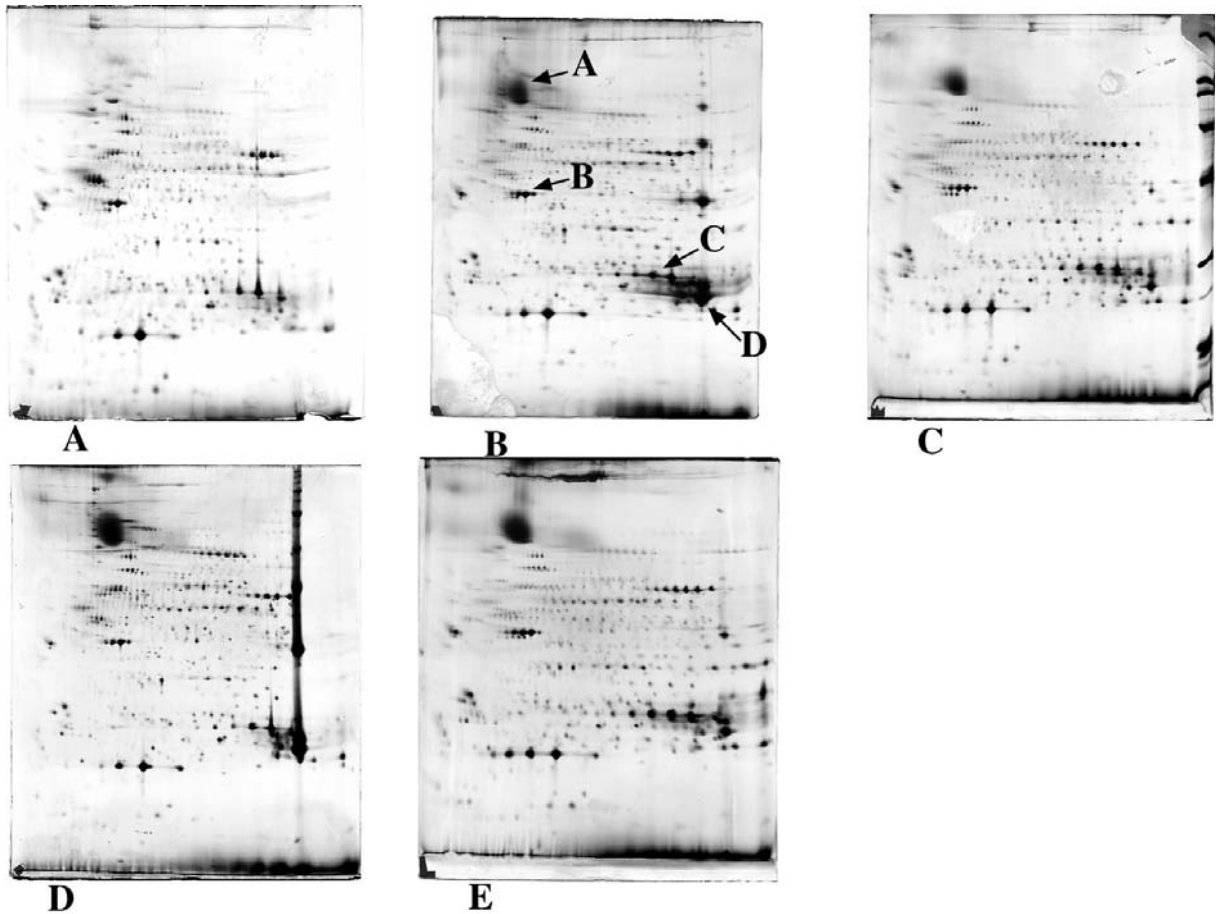

Figure 1. 2D electrophoretic separation of human red blood cells membrane proteins. Red blood cell plasma membrane proteins (150µg) were loaded on the 2D gels. The first dimension is a 4-8 linear pH gradient, and the second dimension a 10% acrylamide gel. Some identified proteins are indicated: A: band 3 (12 transmembrane domains); B: actin; C: stomatin (1 transmembrane domain); D: globin dimer. The proteins are extracted and focused in a solution containing 7M urea, 2M thiourea, 20 mM DTT, 0.4% carrier ampholytes and A: 4% CHAPS; B: 2% Triton X-100; C: 2% Brij56; D: 2% dodecyl maltoside; E: 2% ASB14